# Exciton energy spectra in polyyne chains

Stella Kutrovskaya[1,2,3], Sevak Demirchyan[1,2], Anton Osipov[3,4], Stepen Baryshev[5], Anton Zasedatelev[5,6], Pavlos Lagoudakis[5,6], and Alexey Kavokin[1,2,6,7]

[1] *School of Science, Westlake University, 18 Shilongshan Road, Hangzhou 310024, Zhejiang Province, China*
[2] *Institute of Natural Sciences, Westlake Institute for Advanced Study,*
*18 Shilongshan Road, Hangzhou 310024, Zhejiang Province, China*
[3] *Department of Physics and Applied Mathematics,*
*Stoletov Vladimir State University, 600000 Gorkii street, Vladimir, Russia*
[4] *ILIT RAS Branch of FSRC Crystallography and Photonics" RAS, Shatura 140700, Russia*
[5] *Skolkovo Institute of Science and Technology, 30 Bolshoy Boulevard, bld. 1, 121205 Moscow, Russia,*
[6] *Physics and Astronomy, University of Southampton,*
*Highfield, Southampton, SO171BJ, United Kingdom*
[7] *Spin Optics Laboratory, St. Petersburg State University,*
*198504, Ulianovskaya str. 1, St. Petersburg, Russia*

Recently, we have experimentally observed signatures of sharp exciton peaks in the photoluminescence spectra of bundles of monoatomic carbon chains stabilized by gold nanoparticles and deposited on a glass substrate [1]. Here, we estimate the characteristic energies of excitonic transitions in this complex quasi-one-dimensional nano-system with use of the variational method. We show that the characteristic energy scale for the experimentally observed excitonic fine structure is governed by the interplay between the hopping energy in a Van der Waals quasicrystal formed by parallel carbon chains, the neutral-charged exciton splitting and the positive-negative trion splitting. These three characteristic energies are an order of magnitude lower than the direct exciton binding energy.

*Introduction.*—Being monoatomic chains of carbon atoms carbynes represent ultimate one-dimensional crystals. Carbynes are linear chains of $sp^1$ - hybridized carbon atoms. Their two known allotropes are polyyne, characterized by alternating single and triple electronic bonds between carbon atoms, and cumulene, characterized by double bonds between atoms. Theoretical and recent experimental studies indicate the semiconducting behavior for polyyne chains and quasicrystals, while infinite cumulene chains are expected to be metallic [2, 3]. The strain in finite size chains of polyyne results in the enhancement of the direct band gap, so that experimentally achievable polyyne structures are expected to be emitting visible light in a wide spectral range that is dependent on the specific geometry of the structure [4, 5]. This makes polyyne based nanostructures highly promising for the realization of light-emitting diodes and nano-lasers. In order to predict the quantum efficiency of carbon-based optical nano-emitters one should learn more about excitons in linear carbon chains. The exciton binding energy in $sp^1$ - carbon may be quite large due to the strong two-dimensional quantum confinement. Recently, we detected the strong excitonic features in low-temperature photoluminescence spectra of polyyne bundles stabilized by gold nano-particles and deposited on a glass substrate. The focus of our study was on elongated polyynic molecules, containing straight parts of even numbers of atoms (from 8 to 24) regularly separated with kinks. The synthesised $sp^1$-hybridization chains were packed in hexagon bundles characterised by the distance between neighboring parallel chains of 5.35 A [6]. The chains were hold together by the Van der Waals force (see the schematic in Fig.1(a)). They were grown by the laser ablation in liquid (LAL) and deposited on a fused quartz substrate for the photoluminescence (PL) study. The chains were stabilized by gold anchors attached to their ends [6]. In low temperature PL spectra, we have observed the characteristic triplet structure (Fig. 1(b,c)). The triplet is invariably composed of a sharp intense peak accompanied by two broader satellites shifted by about 15 and 40 meV [1] to the lower energy side of the main peak, respectively. Very interestingly, the triplet structure is found to be nearly identical in carbon chains of different lengths. It moved as a whole with the band-gap variation as the length of the chain changed. We assign the observed sharp peaks to the optical transitions associated with neutral and charged excitons in polyyne bundles. Indeed, as any direct bandgap semiconductor, polyyne is expected to sustain excitons. Sharp resonances that emerge at low temperatures are clear signatures of the excitonic emission. Our time- resolved photoluminescence (TRPL) measurements confirm this assumption [1]. In Ref. [1] we attributed the main peak of every triplet to the neutral exciton transition and two lower energy satellites to charged exciton (trion) transitions, respectively. The main argument supporting this interpretation was the detected dipole polarization of about a half of carbon-metal nanocomplexes that was revealed by their alignment in the presence of the external electric field [6].

To gain better understanding of the observed resonances, one needs to compare the anticipated exciton and trion energies in the system, estimate the thermal hopping energy and the splitting between positively and negatively charged trions. Here we attempt analysing the energy spectra of excitons in carbyne-based nano-sponding to negatively ($X^-$) and positively ($X^+$) charged exciton complexes, trions, consisting of two electrons and one hole and two holes and an electron, respectively, have been studies in doped semiconductor structures [12, 13]. From the theoretical point of view, the excitonic problem is usually treated in the framework of the effective mass approximation. Resolving the Schroedinger equation for the relative motion of electron and hole [14], one finds the fine structure of excitonic transition that varies from hydrogen-like, in bulk crystals [7], to 3D-quantum box spectra in small nanocrystals [10]. More challenging is the description of trion states, where a variety of many-body effects may come into play [15, 16]. In the present study, we rely on the simplest quasi-analytical approach, that seems to be the best adapted at the stage where very little is known about the electronic and optical properties of the relatively complex hybrid nanostructure under study. We extend the approaches developed in Refs.[14, 17] for the variational calculations of exciton states and Ref. [18] for the analysis of trion states. We assume that electron and hole localization radii in the plane normal to the axis of the cylinder are given, respectively, by the parameters $R_e$ and $R_h$ that are much smaller than the exciton Bohr radius. In this case, the electron and hole confinement problem in the systems with use of the variational method. We rely on the effective mass approximation. Using this formalism we were able to shed light on the interplay between several characteristic energies that may be responsible for the fine structures observed experimentally, namely, the positive-negative trion energy splitting, the neutral-charged exciton energy splittings and the hopping energy that splits spatially direct and indirect exciton states.

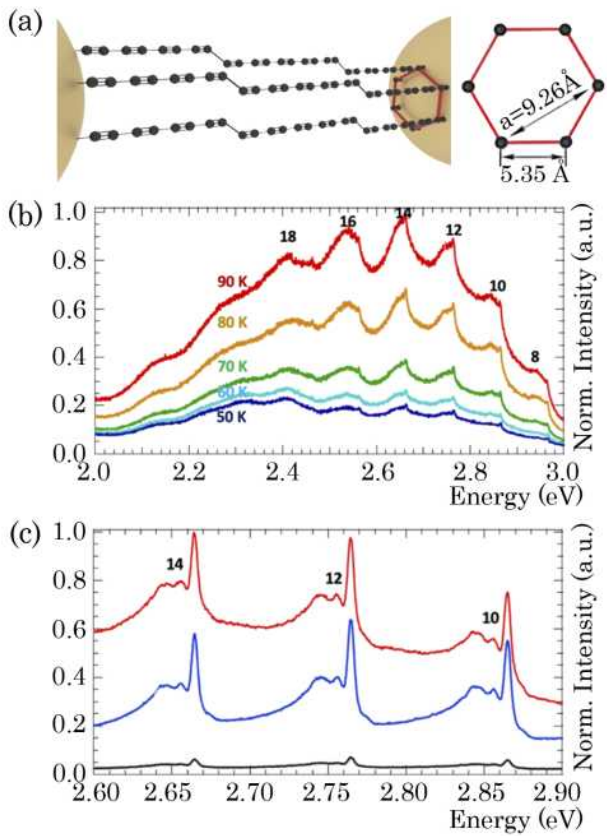

FIG. 1. The top panel (a) illustrates the concept of a stabilized bungle of polyyne chains endcapped by gold nanoparticles (shown as yellow spheres). It also shows the cross-section of the structure that represents a hexagonal Van der Waals crystal. (b) and (c) show the PL spectra of the deposited polyyne chains of different lengths (the number of atoms in the chain is indicated on the top of the corresponding spectral resonance). (b) shows the PL spectra taken at temperatures between 90 to 50K, (c) shows the PL spectra taken at 4K at different excitation wavelength. Red, blue and black curves correspond to the excitation wavelengths of 390, 380 and 370 nm, respectively. The spectra are reproduced from Ref. [1].

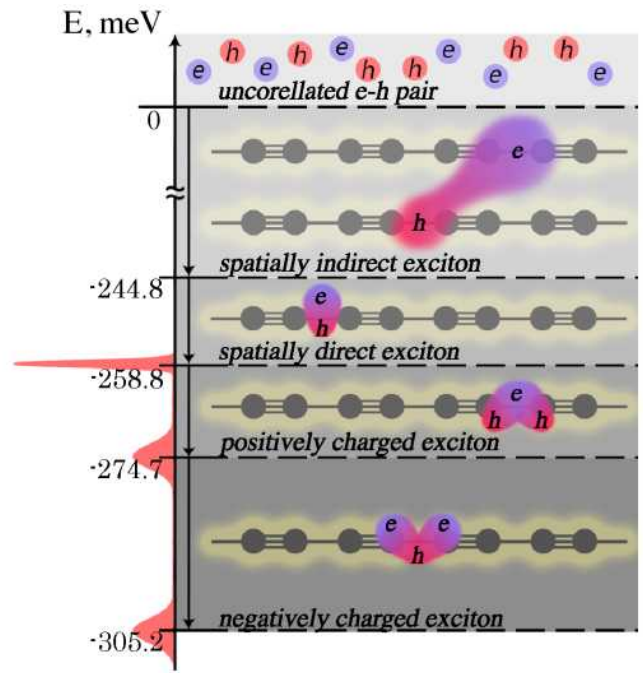

FIG. 2. The schematic energy diagram explaining the fine structure of the excitonic triplet in polyyne chains. Photoexcited electrons and holes may form spatially indirect excitons that do not emit light, as well as charged and neutral spatially direct excitons. The calculated transition energies of direct and indirect excitons as well as $X+$ and $X-$ trions are indicated with respect to the transition energy between uncorrelated electron and hole states.

*The exciton modelling.*—We recall that a Wannier-Mott exciton is a neutral quasiparticle [7], whose optical features strongly depend on the dimensionality of a semiconductor crystal [8]. Much attention has been focused on studies of excitons in strongly confined quantum systems [9, 10]. Stronger confinements typically result in larger binding energies of excitons [11]. In addition to the exciton features, similar spectral resonances corre-normal to the cylinder plane can be decoupled from the Schröedinger equation for the electron-hole relative motion along the axis of the cylinder. The finite size of the cylinder cross-section matters as it is responsible for a finite value of the binding energy of a quasi-one dimensional exciton [19]. We recall that in a purely one-dimensional limit the energy of a ground state of the Coulomb potential tends to minus infinity [20]. We shall work with the exciton trial wave-function taken in form:

$$X(\vec{\rho_h}, \vec{\rho_e}, z) = \Xi_h(\vec{\rho_h})\Xi_e(\vec{\rho_e})\Psi(z). \quad (1)$$

Here $\Xi_{h(e)}$ are hole (electron) wave functions in normal to the axis of the cylinder plane, $\Psi(z)$ is the wave-function of the electron-hole relative motion along the axis of the cylinder. We multiply the exciton Schröedinger equation by the conjugate of the envelopes of the electron and hole wave functions and integrate over the corresponding coordinates. This enables us to derive the Schröedinger equation for an electron-hole relative motion function :

$$H\Psi(z) = E\Psi(z), \quad (2)$$

that contains a smoothed Coulomb potential:

$$H = -\frac{\hbar^2}{2\mu}\nabla_z^2 - \frac{e^2}{4\pi\varepsilon\varepsilon_0}\iint \frac{|\Xi_h(\vec{\rho_h})|^2|\Xi_e(\vec{\rho_e})|^2}{\sqrt{z^2 + (\vec{\rho_h} - \vec{\rho_e})^2}}d\vec{\rho_h}d\vec{\rho_e}, \quad (3)$$

where $\mu = \frac{m_e m_h}{m_e + m_h}$ is a reduced mass. The constituents of the exciton wave function are normalized to unity:

$$\int_{-\infty}^{\infty}|\Psi(z)|^2 dz = \int_0^{\infty}\int_0^{2\pi}|\Xi_{h,e}(\vec{\rho}_{h,e})|^2 \rho_{h,e}d\rho_{h,e}d\phi_{h,e} = 1. \quad (4)$$

For hole and electron envelopes we assume:

$$\Xi_{h,e}(\vec{\rho}_{h,e}) = \frac{1}{\sqrt{\pi}R_{h,e}}\exp\left(-\frac{\rho_{h,e}^2}{2R_{h,e}^2}\right). \quad (5)$$

Within these approximations the Schröedinger equation for the wave function of electron-hole relative motion becomes:

$$\left(-\frac{\hbar^2}{2\mu}\frac{\partial^2}{\partial z^2} - \frac{e^2}{4\pi^3\varepsilon_1\varepsilon_0 R_h^2 R_e^2}\times\right.$$
$$\left.\times \iint \frac{\exp\left(-\frac{\rho_h^2}{2R_h^2}\right)\exp\left(-\frac{\rho_e^2}{2R_e^2}\right)d\vec{\rho_h}d\vec{\rho_e}}{\sqrt{z^2 + \rho_e^2 + \rho_h^2 - 2\rho_h\rho_e\cos(\phi_e - \phi_h)}}\right)\Psi(z) =$$
$$= E\Psi(z), \quad (6)$$

where $\varepsilon_1$ is a dielectric constant corresponding to the spatially direct exciton confined inside the carbon chain. We solve Eq.(2) for $\Psi(z)$ by the variational method. The exciton energy minimization is carried out using the hydrongen-like trial function with a single variational parameter $a$, that has a meaning of the inverse Bohr-radius:

$\Psi(z) = \sqrt{a}e^{-a|z|}$. We find $a$ by minimizing the energy functional $J(a) = \int_{-\infty}^{\infty}\Psi^*(z)H\Psi(z)dz$ that yields:

$$\frac{\hbar^2 a}{\mu} = \frac{e^2}{4\pi^3\varepsilon_1\varepsilon_0 R_h^2 R_e^2}\int_{-\infty}^{\infty}(1 - 2a|z|)\exp(-2a|z|)\times$$
$$\times \iint \frac{\exp\left(-\frac{\rho_h^2}{R_h^2}\right)\exp\left(-\frac{\rho_e^2}{R_e^2}\right)d\vec{\rho_h}d\vec{\rho_e}}{\sqrt{z^2 + \rho_h^2 + \rho_e^2 - 2\rho_h\rho_e\cos(\phi_e - \phi_h)}}dz. \quad (7)$$

For a spatially indirect exciton formed by an electron and a hole confined in neighboring cylinders we obtain:

$$\frac{\hbar^2 a}{\mu} = \frac{e^2}{4\pi^3\varepsilon_2\varepsilon_0 R_h^2 R_e^2}\int_{-\infty}^{\infty}(1 - 2a|z|)\exp(-2a|z|)\times$$
$$\times \iint \frac{\exp\left(-\frac{\rho_h^2 + L^2 - 2\rho_h L\sin\phi_h}{R_h^2}\right)\exp\left(-\frac{\rho_e^2}{R_e^2}\right)d\vec{\rho_h}d\vec{\rho_e}}{\sqrt{z^2 + \rho_h^2 + \rho_e^2 - 2\rho_h\rho_e\cos(\phi_e - \phi_h)}}dz, \quad (8)$$

where $L$ is the distance between the neighboring polyyne chains.

Next, we estimate the binding energy of positively and negatively charged trions. We represent the trial functions for positively and negatively charged trions in the form:

$$X^{\pm}(\vec{r}_{h1,e1}, \vec{r}_{h2,e2}, \vec{r}_{e,h}) = \Xi_{h1,e1}(\vec{\rho}_{h1,e1})\times$$
$$\times \Xi_{h2,e2}(\vec{\rho}_{h2,e2})\Xi_{e,h}(\vec{\rho}_{e,h})\Psi^{\pm}(z_{h1,e1}, z_{h2,e2}, z_{e,h}). \quad (9)$$

The terms composing the trion wave functions are also normalized to unity and the electron and hole envelopes are assumed to be Gaussian isotropic functions, as before. The Schröedinger equations for $X^+$ and $X^-$ states can be written as:

$$\left(-\frac{\hbar^2}{2m_{h,e}}\frac{\partial^2}{\partial z_{h1,e1}^2} - \frac{\hbar^2}{2m_{h,e}}\frac{\partial^2}{\partial z_{h2,e2}^2} - \frac{\hbar^2}{2m_{e,h}}\frac{\partial^2}{\partial z_{e,h}^2} +\right.$$
$$+ V(z_{h1,e1}, z_{h2,e2}) - V(z_{h1,e1}, z_{e,h}) - V(z_{h2,e2}, z_{e,h})\bigg)\times$$
$$\times \Psi^{\pm}(z_{h1,e1}, z_{h2,e2}, z_{e,h}) = E^{\pm}\Psi^{\pm}(z_{h1,e1}, z_{h2,e2}, z_{e,h}), \quad (10)$$

where

$$V(z_\alpha, z_\beta) = \frac{e^2}{4\pi^3\varepsilon_1\varepsilon_0 R_\alpha^2 R_\beta^2}\times$$
$$\times \iint \frac{\exp\left(-\frac{\rho_\alpha^2}{R_\alpha^2}\right)\exp\left(-\frac{\rho_\beta^2}{R_\beta^2}\right)d\vec{\rho_\beta}d\vec{\rho_\alpha}}{\sqrt{(z_\alpha - z_\beta)^2 + (\vec{\rho_\alpha} - \vec{\rho_\beta})^2}} \quad (11)$$

is the effective one-dimensional interaction potential. We assume that $R_{h1} = R_{h2} = R_h$ and $R_{e1} = R_{e2} = R_e$. It is convenient to introduce new coordinates of the relative motion. For $X^+$:

$$z_1 = z_{h1} - z_e,\ z_2 = z_{h2} - z_e,\ Z = \frac{z_{h1} + z_{h2} + \frac{m_e}{m_h} z_e}{\frac{m_e}{m_h} + 2},$$

The same for $X^-$ replacing $h \to e$ and $e \to h$. Now we can separate the motion of a center of mass of a trion and the relative motion of particles and obtain the wave function of the relative motion of $X^+$ and $X^-$ trions:

$$\left( -\frac{\hbar^2}{2\mu} \frac{\partial^2}{\partial z_1^2} - \frac{\hbar^2}{2\mu} \frac{\partial^2}{\partial z_2^2} - \frac{\hbar^2}{m_{e,h}} \frac{\partial}{\partial z_1} \frac{\partial}{\partial z_2} + U^\pm(z_1, z_2) \right) \Psi^\pm(z_1, z_2) = E^\pm \Psi^\pm(z_1, z_2). \quad (12)$$

Here the potentials of interaction in positively and negatively charged trions are introduced as $U^+(z_1, z_2)$ and $U^-(z_1, z_2)$, respectively.

In the following we shall focus on $X^+$ and omit the superscript for simplicity. We shall solve the Schröedinger equation by the variational method assuming the trial function for trions in a singlet state having a form:

$$\Psi(z_1, z_2) = A e^{-a|z_1|} e^{-a|z_2|} (1 + c|z_1 - z_2|). \quad (13)$$

This wave-function should be normalized to unity, which yields for $A$:

$$A = \sqrt{\frac{2a^4}{2a^2 + 3ca + 2c^2}}. \quad (14)$$

Now the energy functional to be minimized can be expressed as:

$$J(a, c) = \frac{\hbar^2}{\mu} \frac{2a^4 + ca^3 + 2c^2 a^2}{2a^2 + 3ca + 2c^2} + \frac{\hbar^2}{m_e} \frac{ca^3 - c^2 a^2}{2a^2 + 3ac + 2c^2} +$$
$$+ \frac{2a^4}{2a^2 + 3ca + 2c^2} \int_{-\infty}^{\infty} \int_{-\infty}^{\infty} \exp(-2a|z_1|) \exp(-2a|z_2|) \times$$
$$\times (1 + c|z_1 - z_2|)^2 U(z_1, z_2) dz_1 dz_2. \quad (15)$$

In the case of $X^-$ the functional is the same as before, with $m_h$ replacing $m_e$ in the second term.

*Results and discussion.*—Solving equation (7) with $\varepsilon_1 = 6$, $R_h = 1\text{Å}$ and $R_e = 2.5\text{Å}$, and using the electron and hole effective masses predicted by the ab-initio calculation [1]: $m_e = 0.078 m_0$, $m_e = 0.09 m_0$, where $m_0$ is the free electron mass, we obtain for the spatially direct exciton $a = 0.351 nm^{-1}$ and the binding energy $E_X = 258, 8 meV$. This seems to us a reasonable value that is close to the wellknown exciton binding energies in carbon nanotubes [21, 22]. Note, that this value may be scaled up or down by tuning the effective dielectric constant. We believe that the value used in this calculation $\varepsilon_1 = 6$ is reasonable for a semiconductor material having a bandgap of about 2.5 eV. Solving the equation (8) for spatially indirect excitons we obtain the variational parameter as $a = 0.297 nm^{-1}$ and the exciton binding energy as $E_X = 244, 8 meV$. Here we accounted for the spatial distribution of the indirect exciton wave-function that spans over two parallel chains by using the reduced effective dielectric constant of $\varepsilon_2 = 4$. This value is obtained by weighted averaging of the effective dielectric constant of a single chain and the dielectric constant of vacuum surrounding the chains, $\varepsilon = 1$. For the positively charged trion $X^+$, we obtain $a = 0.297 nm^{-1}, c = 0.298 nm^{-1}$ and the binding energy $E_{X^+} = 274.7 meV$. For the negatively charged trion $X^-$, we obtain $a = 0.297 nm^{-1}, c = 0.095 nm^{-1}$ and the binding energy $E_{X^-} = 305.2 meV$. Let us now compare the calculation results obtained above with the experimental measurements of the excitonic fine structure in polyyne chains presented in Fig.1(c). Using the parameters listed above we find the positive trion $X^+$ peak shifted from the neutral exciton position by 15.9 meV, and the negative trion $X^-$ resonance shifted from the neighboring $X^+$ peak by 30.5 meV as Fig.2 shows schematically. This appears to be in a very good agreement with the experimental data shown in Fig.1(c). One can easily understand why the negatively charged trion binding energy exceeds one of the positively charged trion looking at the trion relative motion wave-functions shown in Figure 3(d,e). Qualitatively, in a neutral exciton, the hole is located closer to the exciton center of mass than the electron. This is why, an extra hole that needs to be added to create a positive trion should be affected by a strong repulsive potential in the vicinity of the exciton center of mass, and it is subject to an attractive potential far from the exciton center of mass. The opposite is true for an extra electron: it is attracted if it is close to the exciton mass center and repelled otherwise. This is why the three-particle system of a negative trion appears to be stronger localised than the three-body system of a positive trion. As a consequence, the binding energy of a negative trion is larger than one of the positive trion, so that the $X^-$ peak is shifted to the lower energy than the $X^+$ peak. Next, from the model calculation, we obtain the energy splitting between spatially direct and indirect exciton states as 14 meV. This value that plays a role of the thermal hopping energy between the chains is governed by the difference in the shape of the Coulomb potential for two excitons as Fig.3(c) shows. The calculated hopping energy is in a good agreement with the experimental data [1]. Indeed, the thermal dissociation of direct excitons leads to the broadening of exciton peaks that is apparent already at the temperatures of 60-90K (Fig. 1(b)). At the room temperature, the time-resolved PL spectra of [1] show the presence of a fast non-radiative decay channel that is most likely associated with thermal hopping of carriers between chains. This channel is fully frozen out at the liquid Helium temperature. From this ensemble of data, one can estimate the thermal hopping energy as 10-15 meV.

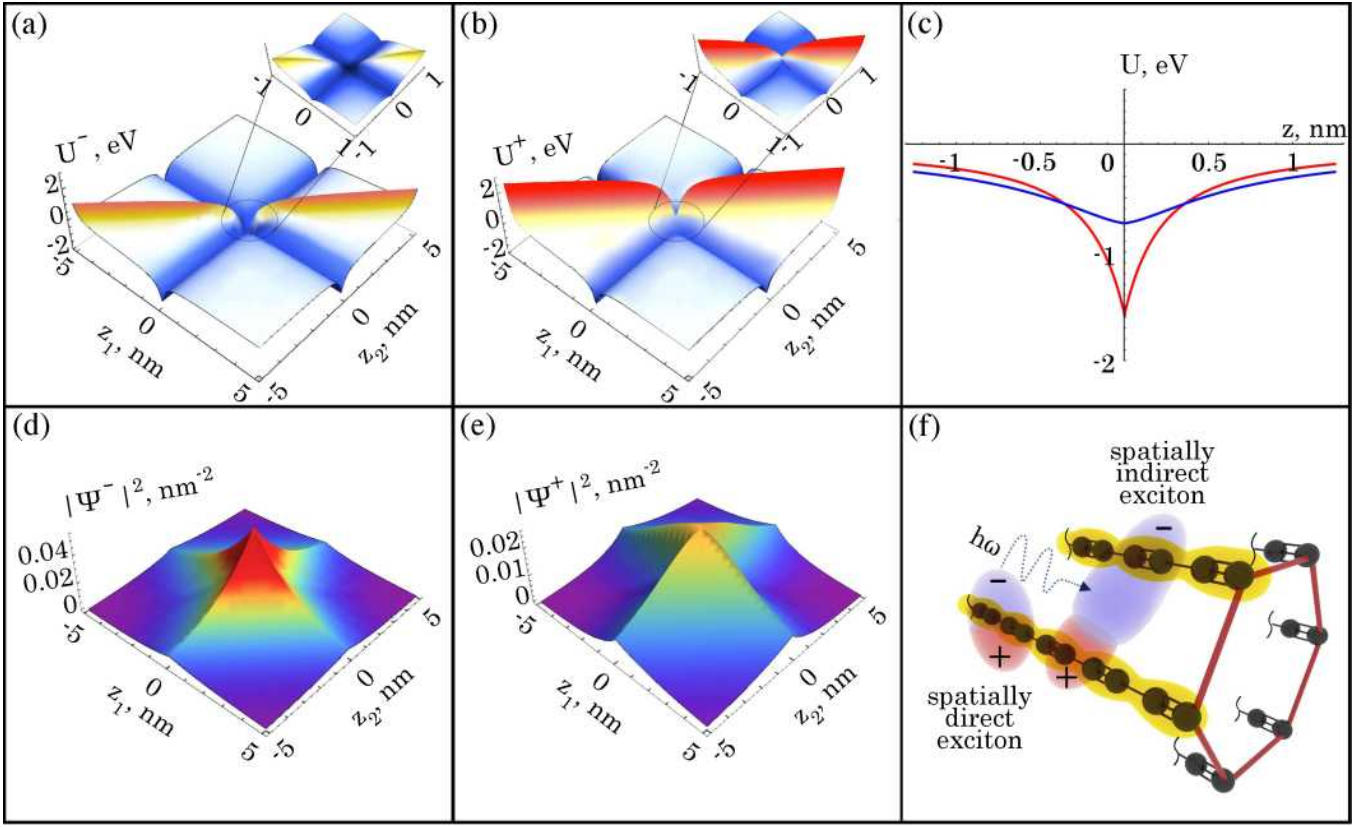

FIG. 3. The charged exciton structure: (a) and (b) show the potentials of Coulomb interaction for the negatively charged trion $U^-(z_i, Z_2)$ and positively charged trion $U^+(z_i, Z_2)$, respectively. (c) shows the calculated interaction potentials corresponding to the exciton states: the red curve shows the Coulomb potential energy of a spatially direct exciton and the blue one shows the Coulomb energy of a spatially indirect exciton. (d) and (e) show negatively charged trion wave-function $T^-(z_i, Z_2)$ and positively charged trion wave-function $T^+(z_i, z_2)$, respectively. (f) shows the schematic illustration of spatially direct and indirect excitons, that may be formed in polyyne polyyne chains bundled in a hexagonal Van der Waals quasicrystal.

*In conclusion,* the developed variational model allowed us to confirm the origin of characteristic energy splittings experimentally observed in low-temperature PL spectra of polyyne chains. We predict the exciton and trion binding energies to be of the order of 250-300 meV. The triplet fine structure that has repeatedly been observed in polyyne chains of different lengths is likely to be due to the spatially direct neutral exciton, position and negative trion peaks, respectively. The non-radiative exciton decay channel observed in the time-resolved PL spectra at the room temperature is most probably associated with the thermal hopping of one of the carriers forming the ex- citon between parallel polyyne chains. Excitons are not destroyed by this hopping process but they become radiatively inactive or dark. We realise that the proposed model has several shortcomings. First, we assumed the carbon chains to be infinitely long. The finite sizes of the chains can be incorporated to the model at the cost of one or two supplementary adjustable parameters. We opted for keeping the model as simple as possible having in mind that, experimentally, the exciton-trion triplet appears to be essentially independent of the length of the chain. Another limitation of the validity of our approach comes from the limited accuracy of the effective mass approximation in nano-systems of such a small size as bundles of polyyne chains composed by 10-20 atoms each. Still, we are confident that the method predicted a correct order of magnitude for the exciton and trion binding and hopping energies, as the comparison with available experimental data certifies. Further experimental and theoretical studies are needed to reveal the spin structure and transport properties of quasiparticles in linear carbon chains.

*Acknowledgement.* The work is supported by the Westlake University, project 041020100118 and the Program 2018R01002 funded by Leading Innovative and Entrepreneur Team Introduction Program of Zhejiang, by RFBR grant 18-32-20006 and by MSHE within the State assignment VlSU 0635-2020-0013. A.K. acknowledges Saint-Petersburg State University for the research grant ID 40847559.